\documentclass[12pt]{iopart}
\usepackage{color}

\usepackage{graphicx}

\begin{document}
\title[Noether symmetry for non-minimally coupled fermion fields]
{Noether symmetry for non-minimally coupled fermion fields\footnote{Dedicated
 to Professor Luis P. Chimento on the occasion of his sixtieth birthday.}}

\author{Rudinei C. de Souza\dag\ and Gilberto M. Kremer\dag
\footnote[3]{To whom correspondence should be addressed
(kremer@fisica.ufpr.br)} }

\address{\dag\ Departamento de F\'\i sica, Universidade Federal do Paran\'a,
 Curitiba, Brazil}

\def\be{\begin{equation}}
\def\ee#1{\label{#1}\end{equation}}
\newcommand{\ben}{\begin{eqnarray}}
\newcommand{\n}{\nonumber}
\newcommand{\een}{\end{eqnarray}}
\newcommand{\lb}{\label}

\begin{abstract}
A cosmological model where a fermion field is non-minimally
coupled with the gravitational field is studied. By applying
Noether symmetry the possible functions for the potential density
of the fermion field and for the coupling are determined.
Cosmological solutions are found  showing that the non-minimally
coupled fermion field behaves as an inflaton describing an
accelerated inflationary scenario, whereas the minimally coupled
fermion field describes a decelerated period,
behaving as a standard matter field.
\end{abstract}



\maketitle

\section{Introduction}

The problems of flatness of the Universe, isotropy of the cosmic
microwave radiation background  and of unwanted relics can be
solved by considering an inflationary era where a scalar field --
the so-called inflaton -- is the responsible for the rapid
accelerated expansion of the primordial Universe  \cite{1}. After
the initial period, the Universe goes into a decelerated era
dominated by  matter fields and the recent astronomical
observations \cite{2} indicates that the Universe has returned to
another accelerated period. The most common theories which explain
the present accelerated era postulates the existence of an exotic
fluid with a negative pressure -- the so-called  dark energy.

Nowadays the search for models that satisfactorily explain the
past and present acceleration periods is  object of intense
investigation. The most popular models are those which consider a
scalar field --  minimally or non-minimally coupled with the
gravitational field --  playing the role of inflaton or dark
energy fields \cite{3}. Recently, modified gravitational theories
were exploited by considering a $f(R)$ term into Einstein-Hilbert
action where the field equations are derived via Noether symmetry
and whose cosmological solutions  can describe  accelerated
periods  of the Universe \cite{4}.

Models which consider fermion fields as sources of gravitational
fields were also investigated in the literature (see e.g.
\cite{5}). Recently, some authors analyzed the possibility that
fermion fields could be the responsible of accelerated  regimes,
playing the role of the inflaton in the primordial Universe or of
the dark energy in the present evolution of the Universe \cite{6}.

Several works in the literature applied Noether symmetry in order
to  search for the forms of couplings and potential densities for
scalar fields (see e.g. \cite{7}), but, to the best of our
knowledge, none have analyzed fermion fields.

The objective of the present work is the study of a general model
of  fermion field coupled to the gravitational field. The search
of the possible forms for the function which is coupled to the
gravitational field and for the fermion field self-interaction
potential density is obtained via Noether symmetry for the
point-like Lagrangian derived from the non-minimally coupled
general model. This approach is very important
because it allows to select the potentials and the couplings
compatible with the symmetry which implies conserved quantities.
Thus the violation of conservation laws is  automatically ruled
out. From this way, Noether symmetry can be seen as a physical
criterion working as a first principle instead of choosing forms
of couplings and potentials without any fundamental
justification.

The model analyzed here describes a spatially flat, homogeneous
and isotropic Universe composed by a fermion field. The evolution
equations of the Universe follow from Einstein's field equations
and Dirac's equation, which are solved for the couplings and
potential densities derived from Noether symmetry. The
cosmological solutions obtained show that (i) a non-minimally
coupled fermion field behaves as an inflaton and can describe an
inflationary period and (ii) a minimally coupled fermion field
acts as a matter field describing a decelerated
phase of the Universe. However, Noether symmetry imposes that
accelerated solutions for the present era are not possible, i.e.,
the fermion field cannot behaves as dark energy.

The work is organized as follows: in Section 2 Einstein and Dirac
equations are derived from a point-like Lagrangian in a spatially
flat Friedman-Robertson-Walker metric, which is obtained from an
action for a fermion field non-minimally coupled to the
gravitational field. The search for the existence of Noether
symmetry for the point-like Lagrangian is the subject of Section
3, where the possible forms of the coupling and of the potential
density are determined.  In  Section 4 the field equations are
solved for  couplings and potential densities found in the
previous section and the cosmological solutions are determined.
The final remarks and conclusions are the subject of the last
section. The signature of the metric adopted is $(+, -, -, -)$ and
natural units $8 \pi G=c=\hbar=1$ are used throughout this work.

\section{Field equations}

The action of a model for a fermion field non-minimally coupled
with the gravitational field reads
 \be
S=\int\sqrt{-g}
d^4x \left\{F(\Psi)R + \frac{\imath}{2} \left[ \overline\psi
\Gamma^\mu D_\mu \psi-(\overline D_\mu \overline\psi) \Gamma^\mu
\psi  \right] - V(\Psi)\right\},
 \ee{1}
where  $\psi$ and $\overline\psi=\psi^\dag\gamma^0$ denote the spinor field and its
adjoint, respectively,  the dagger
represents complex conjugation  and $R$ the Ricci scalar. Furthermore, $V$
is the self-interaction potential density of the fermion field and
$F$ a generic function which describes the coupling of the fermion
and the gravitational fields. In this work it is supposed that
both $V$ and $F$ are only functions of the bilinear
$\Psi=\overline\psi \psi$.

For a spatially flat  Friedmann-Robertson-Walker metric, one can
obtain -- after a partial integration of the action (\ref{1}) --
the point-like Lagrangian
 \be \mathcal{L}=6a
 \dot{a}^2F+6a^2\dot{a}\dot{\Psi}F'+\frac{\imath}{2}a^3(
 \dot{\overline\psi}\gamma^0 \psi -
 \overline\psi\gamma^0\dot{\psi})+a^3V.
 \ee{2}
In the above equation the dot refers to a time derivative whereas the prime to
a derivative with respect to the bilinear $\Psi$ and $a$ denotes
the cosmic scale factor.

From the Euler-Lagrange equation for $\overline\psi$ and $\psi$
applied to the Lagrangian (\ref{2}), it follows  Dirac's equations
for the spinoral field and its adjoint which are coupled with the
gravitational field, namely,
 \be
 \dot{\psi}+\frac{3}{2}H\psi+\imath \gamma^0\psi V'
 -\imath6(\dot{H}+2H^2)\gamma^0\psi F' =0, \ee{3} \be
 \dot{\overline\psi}+\frac{3}{2}H\overline\psi-\imath\overline\psi\gamma^0V'
 +\imath6(\dot{H}+2H^2)\overline\psi\gamma^0F'=0,
 \ee{4}
where $H=\dot{a}/a$ denotes the Hubble parameter.

The acceleration equation follows from the Euler-Lagrange equation
for  $a$, applied to the Lagrangian (\ref{2}), yielding
 \be
 \frac{\ddot{a}}{a}=-\frac{\rho_f+3p_f}{12F}.
 \ee{5}

By imposing that the energy function associated with the
Lagrangian ({\ref{2}) is zero, i.e.,
 \be
 E_\mathcal{L}\equiv\frac{\partial \mathcal{L}}{\partial \dot{a}}
 \dot{a}+\dot{\overline\psi} \frac{\partial
 \mathcal{L}}{\partial\dot{\overline\psi}}+\frac{\partial
 \mathcal{L}}{\partial\dot{\psi}}\dot{\psi}-\mathcal{L}=0,
 \ee{6}
one can obtain Friedmann's equation
 \be
 H^2=\frac{\rho_f}{6F}.
 \ee{7}

The expressions for the energy density $\rho_f$ and for the
pressure $p_f$ of the fermion field are given by \be
\rho_f=V-6HF'\dot{\Psi}, \ee{8} \be
p_f=[V'-6(\dot{H}+2H^2)F']\Psi-V+2(F'\ddot{\Psi}+2HF'\dot{\Psi}+F''\dot{\Psi}^2).
\ee{9}

\section{Noether symmetry}

In terms of the components of the spinor field $\psi=(\psi_1,
\psi_2, \psi_3, \psi_4)^{T}$ and its adjoint $\overline
\psi=(\psi_1^{\dag}, \psi_2^{\dag}, -\psi_3^{\dag},
-\psi_4^{\dag})$, the Lagrangian (\ref{2}) can be
written as
 \be
\mathcal{L}\!=6a\dot{a}^2F+6a^2\dot{a}F'\sum_{i=1}^{4}\epsilon_i\Big(\dot{\psi_i^\dag}\psi_i+\psi_i^\dag
\dot{\psi_i}\Big)
+\frac{\imath}{2}a^3\sum_{i=1}^{4}\Big(\dot{\psi_i^\dag}\psi_i-\psi_i^\dag
\dot{\psi_i}\Big)+a^3V,
 \ee{10}
which is only a function of $(a, \psi_l^\dag, \psi_l, \dot{a}, \dot{\psi_l^\dag}, \dot{\psi_l})$.

Noether's symmetry  is satisfied by the condition
 \be
 L_\textbf{x}\mathcal{L}=0,\qquad \hbox{i.e.,}\qquad
 \textbf{X}\,\mathcal{L}=0.
 \ee{11}
 Above, $\textbf{X}$ is the infinitesimal generator of the symmetry defined by
 \be
 \textbf{X}=C_0\frac{\partial}{\partial
 a}+\dot{C_0}\frac{\partial}{\partial
 \dot{a}}+\sum_{l=1}^{4}\Bigg(C_l\frac{\partial}{\partial
 \psi_l^\dag}+D_l\frac{\partial}{\partial
 \psi_l}+\dot{C_l}\frac{\partial}{\partial
 \dot{\psi_l^\dag}}+\dot{D_l}\frac{\partial}{\partial
 \dot{\psi_l}}\Bigg)
 \ee{12}
 and $L_\textbf{x}$ is Lie's derivative
of $\mathcal{L}$ with respect to the vector $\textbf{X}$ which is
defined in the tangent space. Furthermore, $C_0$, $C_l$ and
${D_l}$ are arbitrary functions of  $(a, \psi_l^\dag, \psi_l)$.

The condition (\ref{11}) when applied to the Lagrangian (\ref{10})
leads to an equation which depends explicitly on $\dot{a}^2$,
$\dot{a}\dot{\psi_l^\dag}$, $\dot{a}\dot{\psi_l}$,
$\dot{\psi_l^\dag}\dot{\psi_n^\dag}$,
$\dot{\psi_l^\dag}\dot{\psi_n}$,  $\dot{\psi_l}\dot{\psi_n}$,
$\dot{a}$,  $\dot{\psi_l^\dag}$ and $\dot{\psi_l}$. By equating
the coefficients of the above terms to zero, one obtains the
following system of coupled differential equations:
 \ben\n
 C_0F+2a{\partial C_0\over \partial
 a}F+a^2F'\sum_{j=1}^4\left({\partial C_j\over \partial
 a}\epsilon_j\psi_j+ {\partial D_j\over \partial
 a}\epsilon_j\psi^\dag_j\right)\\\lb{13}
 +aF'\sum_{j=1}^4(C_j\epsilon_j\psi_j+D_j\epsilon_j\psi^\dag_j)=0,
 \een
 \ben\n
 F'\epsilon_j\psi_j\left(2C_0+a{\partial C_0\over\partial
 a}\right)+aF''\epsilon_j\psi_j\sum_{i=1}^4(C_i\epsilon_i\psi_i+D_i\epsilon_i\psi_i^\dag)
 +aF'D_j\epsilon_j\\\lb{14} +2F{\partial C_0\over
 \partial \psi^\dag_j}+aF'\sum_{i=1}^4\left({\partial C_i\over
 \partial \psi^\dag_j}\epsilon_i\psi_i+ {\partial D_i\over \partial
 \psi^\dag_j}\epsilon_i\psi_i^\dag\right)=0,
 \een
 \ben\n
 F'\epsilon_j\psi_j^\dag\left(2C_0+a{\partial C_0\over \partial
 a}\right)+aF''\epsilon_j\psi_j^\dag\sum_{i=1}^4(C_i\epsilon_i\psi_i+D_i\epsilon_i\psi_i^\dag)
 +aF'C_j\epsilon_j\\\lb{15}
 +2F{\partial C_0\over
 \partial \psi_j}+aF'\sum_{i=1}^4\left({\partial C_i\over \partial
 \psi_j}\epsilon_i\psi_i+ {\partial D_i\over \partial
 \psi_j}\epsilon_i\psi_i^\dag\right)=0,
 \een
 \be
 F'\left({\partial
 C_0\over \partial \psi_j^\dag}\epsilon_i\psi_i+ {\partial C_0\over
 \partial \psi_i^\dag}\epsilon_j\psi_j\right)=0,\qquad
 F'\left({\partial C_0\over \partial \psi_j}\epsilon_i\psi_i^\dag+
 {\partial C_0\over \partial \psi_i}\epsilon_j\psi_j^\dag\right)=0,
 \ee{16}
 \be
 F'\left({\partial C_0\over \partial
 \psi_j}\epsilon_i\psi_i+ {\partial C_0\over \partial
 \psi_i^\dag}\epsilon_j\psi_j^\dag\right)=0,\qquad
 \sum_{j=1}^4\left({\partial C_j\over \partial a}\psi_j- {\partial
 D_j\over \partial a}\psi_j^\dag\right)=0,
 \ee{17}
 \be
 3C_0\psi_j+aD_j+a\sum_{i=1}^4\left({\partial C_i\over \partial
 \psi_j^\dag}\psi_i- {\partial D_i\over \partial
 \psi_j^\dag}\psi_i^\dag\right)=0,
 \ee{18}
 \be
 3C_0\psi_j^\dag+aC_j-a\sum_{i=1}^4\left({\partial C_i\over
 \partial \psi_j}\psi_i- {\partial D_i\over \partial
 \psi_j}\psi_i^\dag\right)=0.
 \ee{19}
 There remains a rest equality which is used for the determination of the potential density,
 namely
 \be
 3C_0V+aV'\sum_{j=1}^4\left(C_j\epsilon_j\psi_j+D_j\epsilon_j\psi_j^\dag\right)=0.
 \ee{20}
 In equations (\ref{13}) through (\ref{20}) it was introduced the  symbol
 $$
 \epsilon_i=\cases{+1\qquad\hbox{for}\qquad i=1,2,\cr
 -1\qquad\hbox{for}\qquad i=3,4.}
 $$

In the following the coupled system of 55 differential equations
(\ref{13}) through (\ref{20}) will be examined. First one infers
from equations (\ref{16}) and (\ref{17})$_1$ that one has two
possibilities  $F'=0$ or $F'\neq0$. Let us analyze the two cases
separately.

\subsection{Case $F'=0$}

If $F'=0$ it follows that $F=$ constant and equations (\ref{16})
and (\ref{17})$_1$ are identically satisfied. Furthermore, from
equations (\ref{14}) and (\ref{15}) one obtains $C_0=C_0(a)$ so
that
 $C_0$ can be determined from equation (\ref{13}), yielding
 \be
 C_0={k\over a^{1/2}},
 \ee{21}
 where $k$ is a constant. From the remaining coupled equations (\ref{17})$_2$ through (\ref{19}) one can determine
 the other two functions of the infinitesimal generator of symmetry $C_j$ and $D_j$, namely
 \be
 C_j=-{3\over2}k{\psi_j^\dag\over a^{3/2}}+\beta\epsilon_j\psi_j^\dag,\qquad
 D_j=-{3\over2}k{\psi_j\over a^{3/2}}-\beta\epsilon_j\psi_j.
 \ee{22}
 Above, $\beta$  is a constant.

Now from the rest equality (\ref{20}) one obtains that the
potential density is a linear function of the bilinear $\Psi$,
i.e., 
 \be 
 V=\lambda\Psi, 
 \ee{23} 
 where $\lambda$ is a constant.

\subsection{Case $F'\neq0$}

From the equation (\ref{20}) one can write 
 \be
 \sum_{j=1}^4\left(C_j\epsilon_j\psi_j+D_j\epsilon_j\psi_j^\dag\right)=-3\frac{C_0}{a}\frac{V}{V'},
 \ee{24} 
 whose  differentiation  with respect to $a$ furnishes
 \be
 \sum_{j=1}^4\left({\partial C_j\over \partial a}\epsilon_j\psi_j+
 {\partial D_j\over \partial a}\epsilon_j\psi^\dag_j\right)=3\Bigg(\frac{C_0}{a^2}
 -\frac{1}{a}{\partial C_0\over \partial a}\Bigg)\frac{V}{V'}.
 \ee{25}

Now, the insertion of equations (\ref{24}) and (\ref{25}) into (\ref{13}) and by 
recalling that $F$ and $V$ are only functions of $\Psi$, leads to
 \be 
 \frac{a}{C_0}{\partial C_0\over \partial a}={V'F\over 3VF'-2V'F}=s, 
 \ee{26} 
where $s$ is a constant.

For $F'\neq0$ equations (\ref{16}) and (\ref{17})$_1$ imply also that $C_0=C_0(a)$, then one can determine
$C_0$ from the equation (\ref{26}), yielding 
 \be 
 C_0=ka^s, 
 \ee {27} 
 with $k$ being a constant.

From equations (\ref{17})$_2$ through (\ref{19}) one obtains that
the functions $C_j$ and $D_j$ read 
 \be
 C_j=-{3\over2}k{\psi_j^\dag a^{s-1}}+\beta\epsilon_j\psi_j^\dag,\qquad
 D_j=-{3\over2}k{\psi_j a^{s-1}}-\beta\epsilon_j\psi_j.
 \ee{28}

 From the rest equality (\ref{20}) it follows also that the potential density
 is a linear function of the bilinear $\Psi$ so that its expression is
 given by equation (\ref{23}). Then, from equations (\ref{23}) and (\ref{26}) one concludes  that the coupling
  is given by a power law $F=\alpha\Psi^p$ where $\alpha$ is a constant. The exponent of the power law for the coupling that follows from equations (\ref{14}), (\ref{15}) and (\ref{26}) must satisfy the two relationships below  from which one can determine $p$ and $s$, namely,
 \be
 \cases{3sp=1+2s,\cr 3p=2+s,} \qquad \hbox{which implies}\qquad (s,p)=\cases{(1,1),\cr (-1,1/3).}
 \ee{29}

 Hence, when $F'\neq0$ the admissible solutions according to Noether symmetry are:
 \begin{itemize}
 \item[\bf(a)]
 \be
 \cases{C_0=ka,\qquad
  C_j=-{3\over2}k{\psi_j^\dag }+\beta\epsilon_j\psi_j^\dag,\cr
 D_j=-{3\over2}k{\psi_j }-\beta\epsilon_j\psi_j,\qquad
 F=\alpha\Psi.}
 \ee{30}
 \item[\bf(b)]
 \be
 \cases{C_0={k\over a},\qquad
  C_j=-{3\over2}k{\psi_j^\dag\over a^2 }+\beta\epsilon_j\psi_j^\dag,\cr
 D_j=-{3\over2}k{\psi_j \over a^2}-\beta\epsilon_j\psi_j,\qquad
 F=\alpha\Psi^{1/3}.}
 \ee{31}
 \end{itemize}

\section{Cosmological solutions}

Note that one cannot distinguish physically the potential density $V$ given by 
(\ref{23}) from a massive term of in the action (\ref{1}), since $V$ is linear in the bilinear
$\Psi$. Then one can consider $V = \lambda\Psi \equiv m\Psi$, where $m$
is the mass of the fermion field.  From now on the coefficient of the potential density will be denoted by $m$.
 
Once the coupling $F$ is a known function of the bilinear
$\Psi$, the search for cosmological solutions are an easy task.
Indeed from Dirac's equations (\ref{3}) and (\ref{4}) one can
build an evolution equation for the bilinear which reads 
\be
 \dot\Psi+3H\Psi=0,\qquad \hbox{so that}\qquad \Psi={\Psi_0\over
 a^3}, 
 \ee{28a} 
 where $\Psi_0$ is a constant.

The time evolution of the cosmic scale factor follows from
Friedmann equation (\ref{7}) and will be analyzed below for the
two cases described in the last section.

\subsection{Case $F'=0$}

The choice  $F=$ constant $=1/2$ refers to
a minimally coupling of the fermion and gravitational fields that
follows from the normalization of the action (\ref{1}) in terms of
natural units. Hence,  Friedmann's  equation (\ref{7})
furnishes that the time evolution of the cosmic scale factor reads
 \be 
 a(t)=[K(t-t_0)]^{2/3}, \qquad\hbox{where}\qquad
 K={3\over2}\sqrt{m\Psi_0\over3}. 
 \ee{29a} 
 One can infer from equation (\ref{29a})$_1$ that it describes a decelerated Universe
dominated by a matter field.

The energy density and pressure of the fermion field follows from
(\ref{8}) and (\ref{9}), yielding 
 \be 
 \rho_f={m\Psi_0\over a^3},\qquad p_f=0. 
 \ee{30a} 
Therefore, in this case the fermion
field  behaves as a standard pressureless matter field.

\subsection{Case $F'\neq0$}
Let us begin with the case (a) where $F=\alpha\Psi$ so that the
Friedmann equation (\ref{7}) becomes 
 \be 
 {da\over a}=\sqrt{-{{m\over12\alpha}}}\;dt. 
 \ee{31a} 
 The above equation shows an exponential behavior of the cosmic scale factor which can describe
an inflationary era. Hence, the fermion field can be identified
with the inflaton and the cosmic scale factor is given by 
 \be
 a(t)=\exp\left[{\cal K}(t-t_0)\right],\qquad\hbox{where}\qquad{\cal
 K}=\sqrt{-{{m\over12\alpha}}}. 
 \ee{32a} 
 Here the energy density and the pressure of the fermion field reads 
 \be \rho_f=-{m\Psi_0\over
 2a^3},\qquad p_f=-\rho_f. 
 \ee{33a} 
 By evoking the weak energy condition which dictates that the energy density is a non-negative
quantity, i.e, $\rho_f\geq0$, one infers from equation
(\ref{33a})$_1$ that $\Psi_0<0$. This last
condition imposes that $\alpha<0$, since the coupling
$F=\alpha\Psi$ must be a positive quantity. Note also that the
condition $\alpha<0$ implies that ${\cal K}>0$.  Furthermore, from
equation (\ref{33a})$_2$ one concludes that the pressure of the
fermion field is always negative and proportional to its energy
density.

In this scenario the time evolution of the
bilinear is given by 
 \be 
 \Psi(t)=\Psi_0\exp\left[-3{\cal K}(t-t_0)\right]
 \ee{34a} 
 thanks to equation (\ref{28a})$_2$.
Moreover, the time evolution of the energy density and pressure of
the fermion field read 
 \be 
 \rho_f(t)=6\alpha\Psi_0{\cal K}^2\exp\left[-3{\cal K}(t-t_0)\right]=-p_f(t). 
 \ee{35a}

Although equation (\ref{32a}) predicts an eternal accelerated
expansion for the Universe, equations (\ref{34a}) and (\ref{35a})
show that the source of the accelerated expansion should come to
an end,  since the potential density and the energy density of the
fermion field tend to zero at a finite time.

For the case (b) where $F=\alpha\Psi^{1/3}$ the
Friedmann equation (\ref{7}) does not have a solution.

\section{Final remarks and conclusions}

In this work it was consider a classical 
spinor field which is  understood as a set
of complex-valued space-time functions which transform according to
the Lorentz group. More details about classical spinors can be
found in  the work by Armend\'ariz-Pic\'on  and Greene  in reference \cite{5}. About
such a consideration  one has to observe that: (i) the
spinorial field can be treated classically if its state is close
to the vacuum and (ii) the expectation value of a spinorial field
in a physical state is not a Grassmannian number but a complex
number.

In the literature (see e.g. \cite{5,6}) varied forms of the
potential density and coupling for fermion fields were proposed in
order to describe cosmological models with  accelerated and
decelerate periods of the Universe.
 The results obtained in the present work show that the functions for the
potential density and for the coupling  are very restrictive, once
Noether symmetry is satisfied.

Furthermore, it was also shown that: (i) the non-minimally
coupling of the fermion and gravitational fields leads to an
accelerated expansion describing  an inflationary era where the
fermion field behaves as an inflaton and (ii) a minimally coupled
fermion field implies a decelerated regime where the fermion field
acts as a standard matter field. 

\ack
The authors acknowledge the support from CNPq (Brazil).

\section*{References}



\end{document}